# A RESIDUAL STRESS CHARACTERIZATION METHOD OF A SMALL DIAMETER WIRE BY MATTER REMOVAL

Julien Vaïssette[a], Manuel Paredes[a], Catherine Mabru[a]

[a] *Institut Clément Ader (ICA) ; Université de Toulouse ; CNRS, IMT Mines Albi, INSA, ISAE-SUPAERO, UPS, 3 rue Caroline Aigle, 31400 Toulouse, France*

___

**ABSTRACT**

Standard spring design is based on the assumption that there is no residual stress in the wire. This assumption can be quite strong for small wires that undergo a significant wire drawing during the manufacturing process. The main objective of this work is to design and test a method for characterizing residual stresses in small diameter drawn wires. The basic principle consists in removing a part of the material and then measuring the associated displacements. This material removal was simulated by finite elements and experimented in real situation. The first results are encouraging because they seem to show that the residual stresses in the wire studied are significant, and justify our investigations.

___

## 1. Introduction

The standard NF EN 13906-1 [1], on which many calculations for the design of helical springs are based [2], relies on the assumption of the absence of residual stresses in cold-drawn steel wires. This means that before spring forming, the steel wire is supposed to be completely relaxed of first-order, second-order and third-order residual stress [3]. Consequently, the design standard makes no mention of the presence of residual stresses in spring wires. However, when steel wires enter the forming phase, they come out of a drawing process that is very heavy in plastic deformations, sometimes followed by a heat treatment. It is therefore a strong assumption to assume that the heat treatments following the wire cold-drawing and the spring forming are efficient enough to relax all residual stresses.

Indeed, the residual stress profile in the steel wires at the end of the drawing process has widely been studied [4-5], as for the residual stress profile after spring forming [6]. Concerning the residual stress profile after cold drawing, it is found in every cold-drawn wire: the skin of the wire is loaded with residual tensile stresses, while the core of the wire is stressed in compression.

The assumption of no residual stresses becomes problematic in the design of springs considering the influence that the distribution and intensity of residual stresses can have on the mechanical behavior of springs, especially in fatigue [7]. However, it is difficult to dispense with this hypothesis because of the difficulty encountered by spring manufacturers to characterize the residual stresses in the steel wires they receive. Indeed, if various methods of characterization of residual stresses exist (incremental drilling method [8] and Sachs method

___

\* Corresponding author.  author@institute.xxx*(Times New Roman 12pt)*

[9] for instance), they are difficult to apply on wires with small diameters: for diameters inferior to 2 mm, drilling in the wire is extremely difficult.

In order to always refine the design of small diameter springs, the present work proposes to develop a method to characterize these residual stresses. In this study, it was chosen to confront our method to a wire of small diameter (0.8 mm). Moreover, the steel that composes the wire investigated is an austenitic steel that is common in spring manufacturing: the AISI 302 steel.

## 2. Experimental method

### 2.1. Coating and polishing of the steel wires

This residual stress characterization method is inspired by the incremental drilling method, because it relies on the same principle of matter removal: a layer of material is removed from the steel wire and the displacement induced by this removal is measured. To remove the layer of material, the wire is embedded into resin in order to maintain it during the operation of matter removal. However, it is impossible to use the incremental logic of the incremental drilling method since it is necessary to dissolve the resin in which the wire is embedded to measure the deformation of the wire. Thus, only one layer of matter is removed, and the level of residual stress is identified from this deformed profile.

The experimental protocol is simple: it consists of coating wire samples (in this work, the coated steel wire is 20 mm long because of the size of the coating press), measuring their curvature, polishing the wires longitudinally to remove material, dissolving the resin, and measuring the curvature of the wire after dissolution (see Fig. 1).

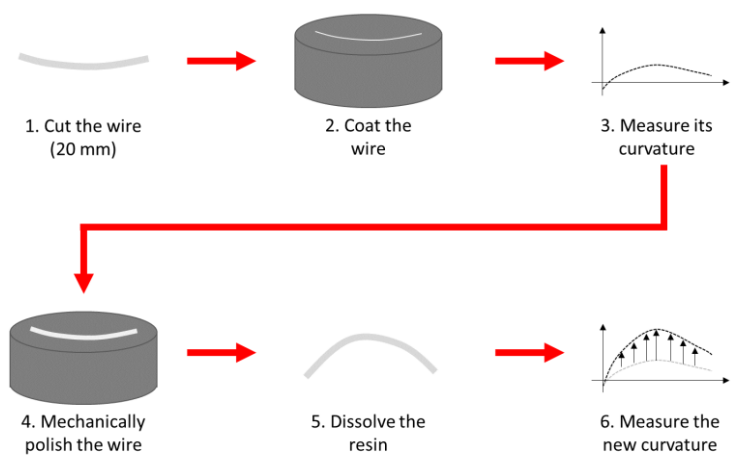

Fig. 1. The steps of the matter removal method.

Behind the simplicity of the steps of this residual stress characterization, there are some critical points. The first critical point is during the first curvature measure. Indeed, curvature measurements are performed on Alicona InfiniteFocusSL, an optical 3D measurement device in this work. This device requires placing the sample in a perfectly stable position, since the stage moves during the measuring process; and the sample needs to be parallel to the optical sensor, as the depth of focus of the measuring device is extremely thin. It is a difficult point because it means that the wire must be carefully positioned during the coating process (performed on a Simplimet 2000 press from Bhueler), to be as parallel to the press plate as possible. If it is not parallel, the measure is less accurate and the layer removal cannot be uniform over the entire length of the wire, making it impossible to interpret the curvature variation of the wire.

Once the wire is parallel to the sample upper face, it is necessary to verify the bottom face of the sample. Indeed, according to the resin used during the coating process, the bottom face of the sample can be curved, leading to instability of the optical measurement, because the curved bottom face is the face in contact with the stage. In this case, it may be necessary to polish the bottom side to make it as flat as possible.

The coating of the wire is done under heat and pressure, at 150°C and 2.9 bars. It was then decided to determine if these coating conditions had an influence on the curvature of the wire. In order to verify this point, the curvatures of 4 wires were measured a first time after the coating and a second time after the dissolution of the resin, without polishing. The results were satisfying: the curvatures measured before and after dissolution are not equal but the difference in curvature is largely negligible compared to the radial displacement caused by the material removal. Indeed, the radial displacement did not exceed 2 to 3 microns over the length of the studied wires against several tens of microns after removal of a layer of material.

After coating the wires and measuring the wire curvature before polishing, the samples are polished on an automatic polishing machine. This step is critical because it is important not remove too much matter to let the wire embedded in the resin, and the mechanical polishing step must avoid any influence on the residual stress profile in the wire. A succession of very short polishing duration with coarse grades (P320, P600 and P1200) are then used and the polishing process is finished by long duration polish (several minutes) with 3 µm and 1 µm diamond suspensions.

The resin used to coat the wires is the acetone soluble resin ESCIL. This reference was chosen since the wires are extremely sensitive to deformations, the removability of the resin is a great feature: the wire falls by itself out of the coating when put in an acetone bath.

When the wire has fallen from the resin sample, the curvature of the wire after matter removal is measured with the same optimal measuring device. Again, the sample must be as stable as possible on the stage. With this mind, the polished wire is positioned on its polished area (see Fig. 2).

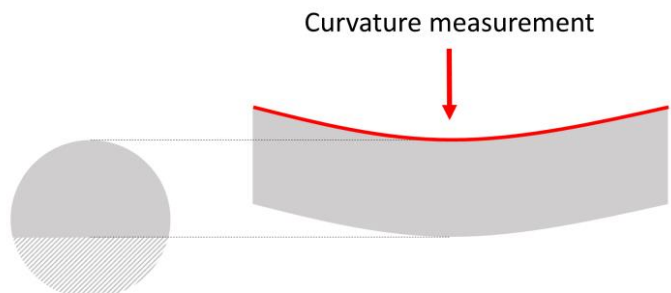

Fig. 2. The curvature measurement is performed on the opposite of the polishing area.

During the calculation of the curvature variation, the first measurements need to be reversed. Indeed, the first curvature measurement (before polishing) is made on one side of the wire, and the second measurement (after polishing) is made on the opposite side. The curvature variation is then equal to the sum of the two measured curvatures, and not the difference between the first and the second curvatures.

To quantify the curvature variation after matter removal, it was found that the easiest method is to concentrate on the radial displacement of the wire. It expresses the variation of curvature without using radius calculations. The presented method therefore provides the radial displacement of the wire caused by matter removal. This radial displacement is caused, as we know, by the removal of a layer of material from the steel wire with residual stresses, which imposes to the wire the search for a new internal equilibrium. As the layer removed from the steel is mostly stressed in tension, the residual stress profile is no longer balanced and the wire bends to find a state of macroscopic equilibrium. The more the wire is containing residual stress, the more it bends after polishing. The proposed method then consists in trying to attribute a residual stress profile to a radial displacement of the wire after removal of a layer.

*2.2. Finite Element simulations*

In order to compare the displacements obtained experimentally, numerical simulations were conducted on Abaqus. The aim is to estimate the residual stresses generated by a severe deformation process comparable to cold drawing, and have an idea of the intensity of the displacement that could occur when a layer of matter of a wire severely deformed is removed. This Finite Element simulation is performed in two steps.

The first step is to reproduce a drawing process numerically. A diameter reduction is simulated on Abaqus (see Fig. 3a), based on the material law studied in the work of C. Levreau [10]. After this simulation a residual stress profile is obtained at the end of the drawing process on a stainless steel (see Fig. 3b). The wire's diameter before simulated cold drawing is 0.86 mm, and after simulated cold drawing, it is 0.84 mm. To simplify the simulation process, the wire before cold drawing is straight, unlike the cold drawn wire that all have a natural curvature.

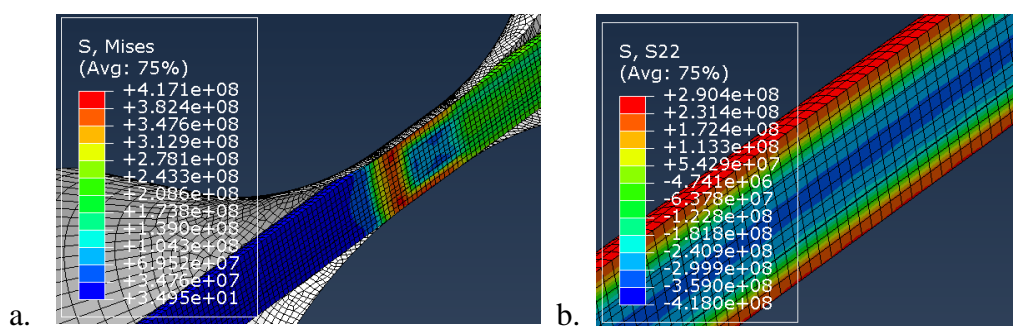

Fig. 3. (a) Von Mises stress during cold drawing in Pa (b) Axial residual stress after cold drawing simulation on the 0.8 mm wire.

The second step is to simulate the matter removal by deactivating some elements of the mesh. With this layer deactivation, the wire is released to see how it deflects by only fixing the end of the wire as boundary condition. By doing so, the wire finds a new macroscopic equilibrium and its curvature changes from a perfectly straight wire to a curved wire. Thanks to this simulated displacement, it is possible to have an idea of the curvature modification generated by layer removal on a residual stressed wire. It is also possible to estimate (within the framework of the starting hypothesis) the radial displacement generated by different depth of layer removal. This simulation was performed with 4 different material removals: 0.1 mm, 0.2 mm, 0.3 mm and 0.37 mm (see Fig. 4).

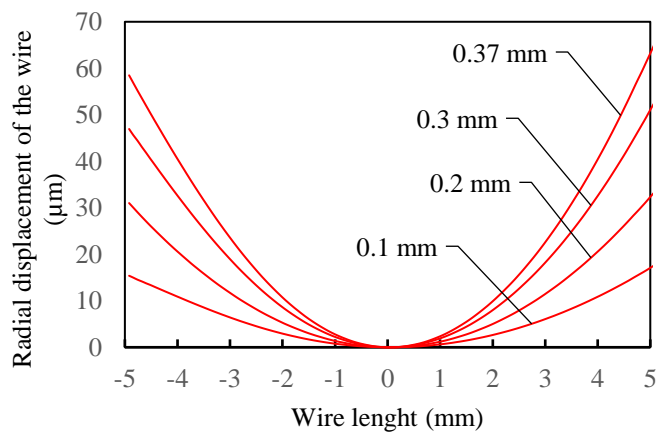

Fig. 4. Radial displacement of a cold-drawn wire after removal of different layer depth.

The results of layer removal simulation show clear radial displacement of the cold-drawn wire. It can also be seen that the greater the depth of matter removal, the greater the radial displacement. In addition, if referring to Fig.3, it is clear that this significant radial displacement reflects a significantly severe residual stress profile because the axial residual stress is up to 260 MPa at the skin of the wire and -324 MPa at the core. It proves that for such residual stress profile in the wire, a measurable radial displacement after matter removal characterization method could be observed.

## 3. Results

This residual stress characterization method was applied on 4 austenitic steel wires AISI 302, whose diameters are 0.8 mm and whose length are 20 mm. The radial displacement obtained by removing respectively 0.1 mm (See Fig. 5a), 0.17 mm (see Fig. 5b), 0.16 mm (see Fig. 5c) and 0.17 mm (see Fig 5d) are substantial, because they exceed the radial displacement that was expected thanks to the numerical simulations.

The results lead to the same conclusion for the 4 tested wires: the radial displacement observed after removal of material is non-negligible, and invariably higher than what the numerical simulations predicted, even though the stress levels are significant. The most impressive result is for the 0.17 mm layer removals: the radial displacement was expected to lie between 15 µm and 30 µm at the furthest measuring point from the middle of the wire. Instead, the radial displacement at this point is superior to 70 µm.

This extreme result is not caused by a statistical error: a remarkable repeatability is found in the radial displacement measurements. Indeed, three of the four wires were removed by the same thickness of material (0.16 mm to 0.17 mm) and their radial displacements are extremely close to each other (see Fig. 6). This suggests the radial displacement measuring method is reliable.

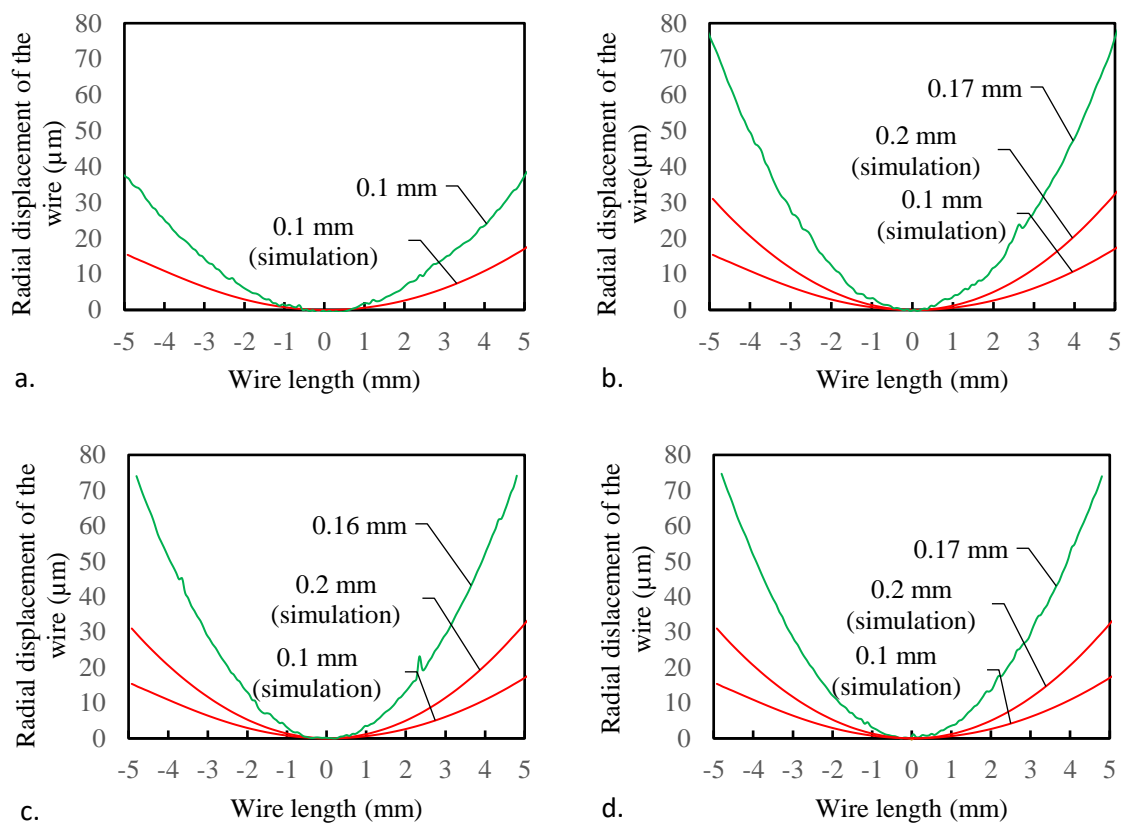

Fig. 5. (a) Radial displacement of the wire after the removal of a 0.1 mm layer (b) After the removal of a 0.17 mm layer (c) After the removal of a 0.16 mm layer (d) After the removal of a 0.17 mm layer.

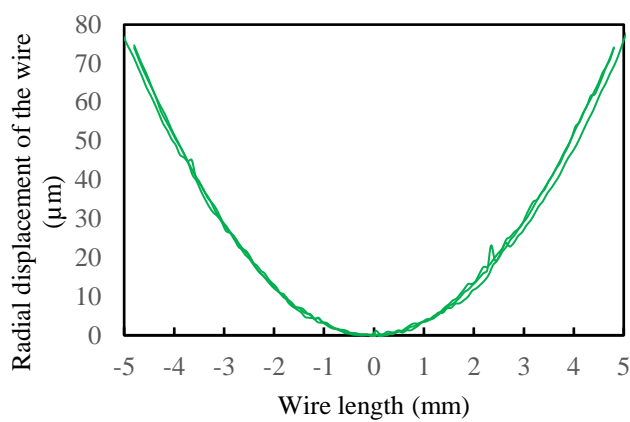

Fig. 6. Radial displacement of three wires with the same thickness of layer removed (0.16 mm to 0.17 mm).

At this point, it is clear that the assumption of no residual stresses in the steel wire before forming is far from the reality of the studied wire. Indeed, the removal of material has led to a radial displacement of the wires that leaves no doubt about the presence of residual stresses. On the other hand, the numerical model underestimates the displacement caused by the removal of material. This implies that this numerical model seems to underestimate the level of residual stresses.

In order to confirm the shape of the residual stress profile obtained with the matter removal model, it was decided to run micro-indentation tests and observe the radial evolution of the hardness in the wire. As the residual stress level influences hardness, these measures allow underlining the presence of a potential residual stress variation in the wire.

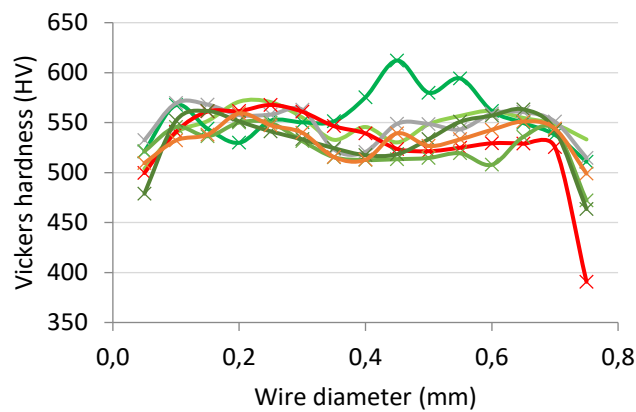

Fig. 7. Vickers hardness measured with micro-indentation HV0.3 along the diameter, on 7 wires AISI 302, diameter 0.8 mm.

In Fig. 7, a clear variation of hardness along the diameter is visible. It is due to measured points that were close to the skin (0.05 mm from the skin). Going this close to the outer layer of the wire, the tensile stress effects can be seen on the hardness level. This allows estimating the depth of the layer subjected to tensile stress: it seems to be around 0.05 mm deep.

Moreover, it is also possible to have a first idea of the magnitude of the residual stress profile. Indeed, a previous study [11] has proposed a relationship between Vickers hardness and axial residual stress on another austenitic steel (AISI 316). With the assumption of a similar behavior between AISI 302 and 316, it is possible to roughly estimate the magnitude of residual stress (see Fig.8).

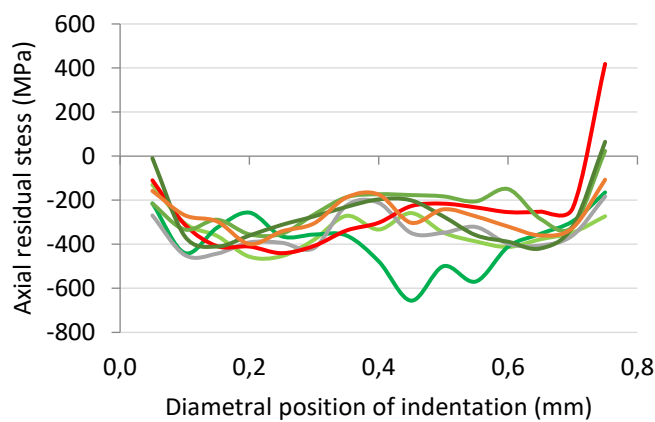

Fig. 8 The evolution of the estimated residual stress values along the diameter of the wire.

This figure confirms that there must be residual stress in cold-drawn wires, as observed numerically and experimentally by matter removal. However, it can be seen that the shape of the residual stress profile is not exactly the same as the shape obtained by numerical simulations: the compressive residual stress seem to be extremely flat in the core of the wire and on a wide radial distance. And the inversion of the residual stress values seem to be a lot sharper than what numerical simulations estimate.

## 4. Discussion

In Fig. 3, it can be seen that the numerical model evaluates the axial residual stresses at high magnitudes. More precisely, the numerical model estimates radial stress to be 260 MPa at the skin of the wire and -324 MPa at the core. These values, which are not negligible, seem to be lower than those actually found in the steel wire at the end of the drawing process that was studied, because they lead to a lower radial displacement of the wire after matter removal.

However, it is difficult to clearly assess the relevance of the proposed methodology to precisely determine the level of residual stresses in the wires of small diameter. The impact of each step of the proposed protocol on the robustness of our model has been checked. It was established that the temperature and pressure of the coating process did not change the residual stress profile of the wires, and that the measurements were repeatable. But at that time, it is not possible to guarantee that the mechanical polishing used during the matter removal step does not add parasitic residual stress in the wires. Previous works have shown that mechanical polishing can modify the residual stress level [12-13]. Moreover, these works show that the parasitic residual stress added is compressive stress. This implies that mechanical polishing can increase the curvature variation, since the wire core is already subjected to compressive stress and the removed skin is subjected to tensile stress. This increase in curvature can lead to overestimate the amount of residual stress present in the wires tested. The stresses measured in previous work [13-14] exceed 50 MPa (more than a tenth of the stresses obtained in the simulation). The altered depth is not mentioned in these works. It is therefore necessary to estimate the magnitude of the compressive residual stresses that are added during the mechanical polishing step. It is also necessary to estimate the depth of the layer affected by this additional compressive stress as well. This is mandatory in order to integrate the influence of polishing in the proposed model. Works are currently in progress in this aim.

## 5. Conclusion

This paper presents an original experimental method for qualitatively characterizing residual stresses adapted to the case of small wires (diameter less than 1 mm). The principle consists in coating the wire in a soluble resin and then gently polishing the sample in order to make a flat cut parallel to the wire axis. As for the other matter removal residual characterization method, the objective is to observe the deformation generated by the release of residual stress after the layer is polished. In order to estimate the magnitude of residual stress generating the curvature change, a finite element simulation of the entire protocol was used.

Experiments conducted on a 0.8 mm diameter AISI 302 wire showed a very good repeatability and highlighted a very high intensity of residual stresses, which justifies the need to properly characterize the residual stress profile of cold-drawn wires. However, this approach still needs to be pursued in order to refine the relationship between the profile obtained after polishing and the residual stresses in the wire in order to reach a quantitative measure.